\documentclass[11pt]{article}
\usepackage{moriond,epsfig}

\def\Journal#1#2#3#4{{#1} {\bf #2}, #3 (#4)}

\def\PLB{{\em Phys. Lett.}  B}
\def\PRL{\em Phys. Rev. Lett.}
\def\PRD{{\em Phys. Rev.} D}

\def\be{\begin{equation}}
\def\ee{\end{equation}}
\def\bea{\begin{eqnarray}}
\def\eea{\end{eqnarray}}

\begin{document}
\vspace*{4cm}
\title{Solar Neutrino Results from Super-Kamiokande}

\author{ M. B. Smy\\
for the {\bf Super-Kamiokande Collaboration}}

\address{Department of Physics and Astronomy, 4182 Frederick Reines Hall,\\
UC Irvine, California, USA}

\maketitle\abstracts{
Super-Kamiokande has measured the solar neutrino flux
using elastic neutrino-electron scattering in water. The
measured flux is 
$2.32\pm0.03$(stat)$^{+0.08}_{-0.07}$(syst)$\times10^6$/(cm$^2$s)
based on the energy range of 5 to 20 MeV for the recoiling electron.
The time-dependence
and energy spectrum of the recoiling electrons were studied to
search for two-neutrino oscillation signatures. The absence of either
significant zenith angle flux variation or distortions of the
recoil electron spectrum places strong constraints 
on neutrino mass difference and mixing.
In combination with the flux measurement, two allowed regions at large
mixing are found.}

\section{Introduction}
Using various detection methods, measurements of the solar neutrino
flux~\cite{homestake,kam,gallex,sage,sk} fall short of the flux
predicted by the standard solar model~\cite{bp2000} (SSM). This discrepancy
is known as the ``solar neutrino problem''. 
Neutrino flavor oscillations, similar to those seen in
atmospheric neutrinos~\cite{skatmos}, are a
natural explanation for the discrepancy.
In addition to a conversion in vacuum,
a matter-induced resonance in
the sun~\cite{msw} may sufficiently enhance the disappearance
probability of solar neutrinos even for
small neutrino mixing. Matter effects can also modify the
survival probability for neutrinos which pass through the
earth~\cite{earthmatter}.

\begin{figure}[tbh]
\psfig{figure=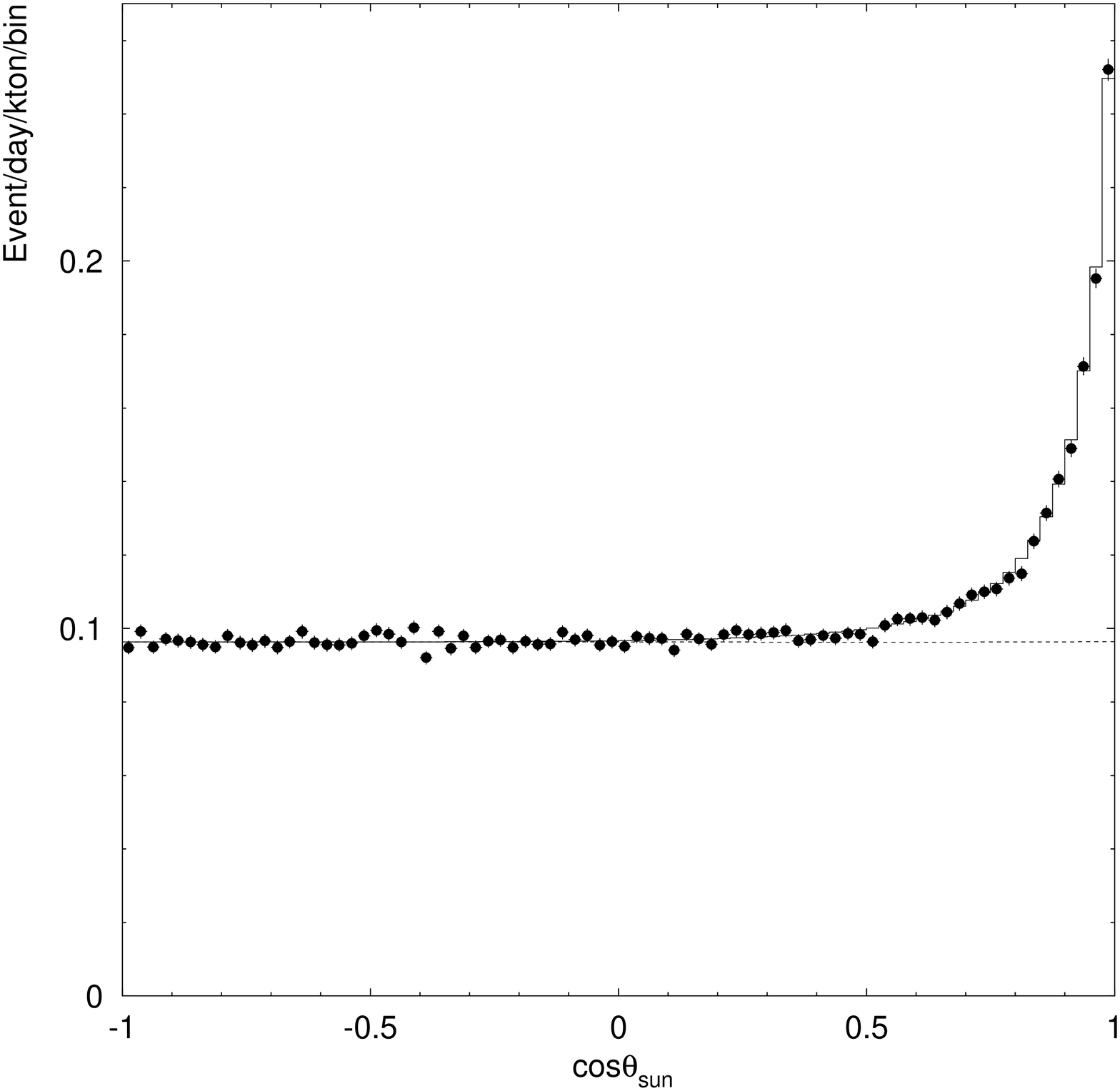,width=8.2cm}
\psfig{figure=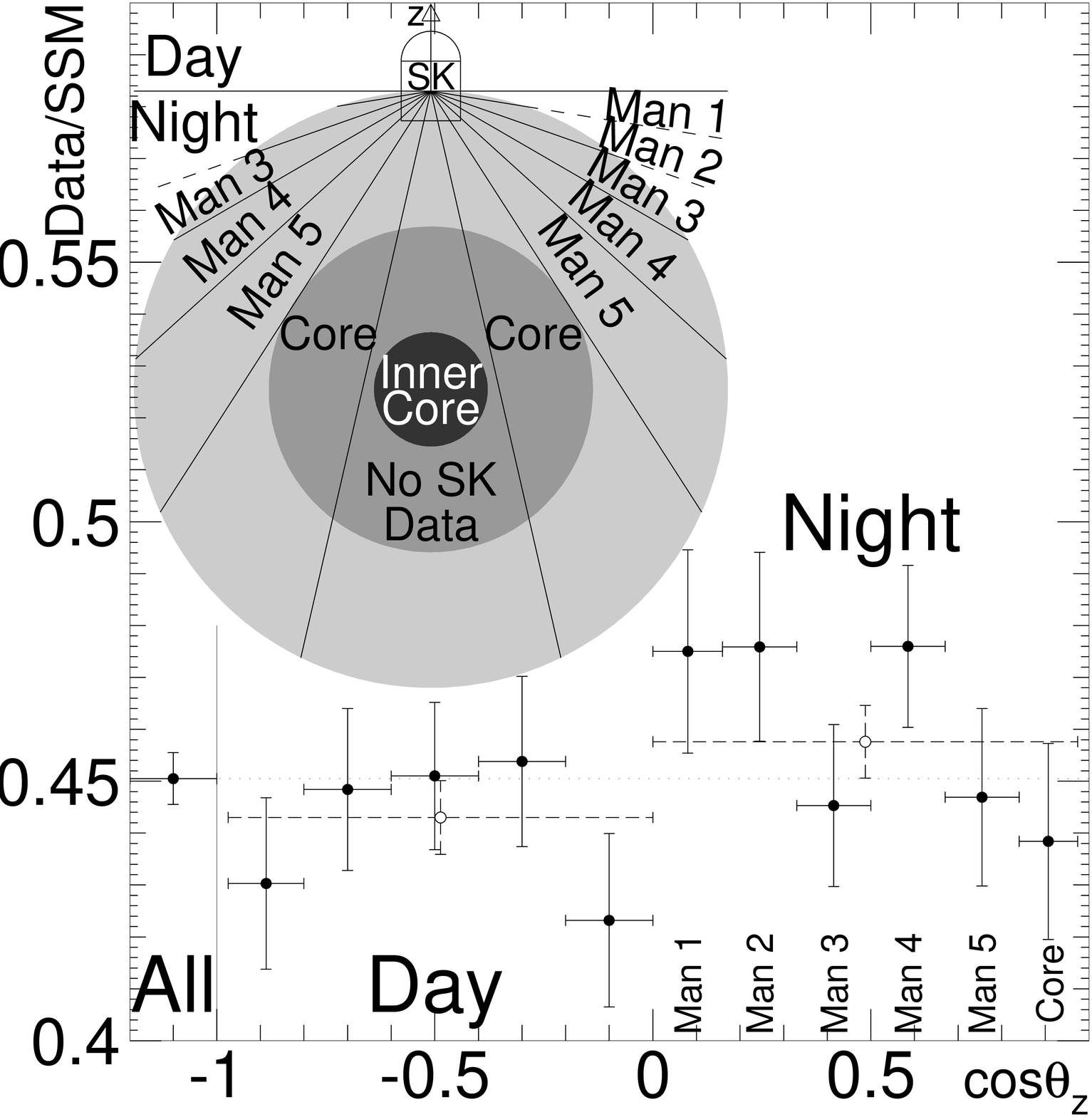,width=7.8cm}
\caption{Angular distribution of solar neutrino
event candidates (left) and solar neutrino flux
in units of the flux expected from the
BP2000~\protect\cite{bp2000} model
(right). The strongly forward-peaked cross
section of elastic neutrino-electron scattering
leads to a clear peak of events pointing back
to the sun. The recoil
electron energy is limited between 5 and 20 MeV.
Next to the total flux (45.1\% BP2000; dotted
line), the flux based on a
day-time (night-time) sub-sample is shown
(open circles; dashed line).
The day-sample is further subdivided into five zenith
angle bins, the night-sample into six (solid circles).
Neutrinos in the last zenith angle bin pass through
the core of the earth as indicated in the sketch
of the earth's structure (right).
\label{fig:zendist}}
\end{figure}

Super-Kamiokande (SK) is
a 50,000 metric ton water Cherenkov detector. Solar Neutrinos are
detected via elastic neutrino-electron scattering if the reconstructed
energy of the recoiling electron is above 5 MeV. Vertex, direction and
energy are reconstructed using the timing and pattern of the Cherenkov
light produced by the recoil electron. About 11,000 photomultiplier tubes
(PMTs) view the ``inner detector'', a cylindrical volume containing 32,000
metric tons of water. Restricting the vertex of the solar neutrino
event candidates to further than 2m from the PMTs leaves 22,500 tons of
fiducial mass.
Due to the energy threshold of 5 MeV,
only the $^8$B decay and the He---proton reaction ({\it hep}) branches
of the solar
neutrino spectrum are accessible.
To extract the solar neutrino flux from the sample of solar neutrino
event candidates, the reconstructed direction of the recoiling
electron is used (see~\cite{sk} for details).
The angle $\theta_{\mbox{sun}}$ is defined as
the angle between the reconstructed electron direction and the
vector pointing from the sun to SK.
In figure~\ref{fig:zendist}, the distribution of cos$(\theta_{\mbox{sun}})$
is shown. The  strong
correlation of the recoil electron direction with the neutrino direction
leads to a clear peak at cos$(\theta_{\mbox{sun}})$=1.
Using 1258 days of data, the measured flux is (see figure~\ref{fig:zendist})
\be
\phi=2.32\pm0.03\mbox{(stat)}^{+0.08}_{-0.07}\mbox{(syst)}\times
\frac{10^6}{\mbox{cm}^2\mbox{s}}
\ee
or
$45.1\pm0.5$(stat)$^{+1.6}_{-1.4}$(syst)\% of the flux
predicted by the BP2000~\cite{bp2000} SSM.

The solar neutrino flux from the {\it hep} branch is
expected to be about three orders of magnitude
smaller than the $^8$B branch. The endpoint of the
{\it hep} spectrum (18.77 MeV) is higher than the $^8$B
spectrum endpoint (16 MeV). If the
${\it hep}$ flux differs from expectation, the recoil electron spectrum
will appear distorted. To place an upper limit on the
${\it hep}$ flux, we define an energy range by maximizing
sensitivity to {\it hep} neutrinos. The energy range chosen
is 18 to 21 MeV. In this range we observe 1.3$\pm$2.0
solar neutrino events, or less than 3.9 events at 90\%
confidence level. From BP2000 ($\phi_{hep}=9.3\cdot10^3/($cm$^2$s))
we expect 0.9 {\it hep} neutrinos, so 3.9 events correspond
to an upper limit of the {\it hep} flux of $40\cdot10^3/($cm$^2$s)
at 90\% confidence level (4.3 times the BP2000 prediction assuming
no oscillation).

\section{Time Variation of the Flux}
\begin{figure}[tbh]
\psfig{figure=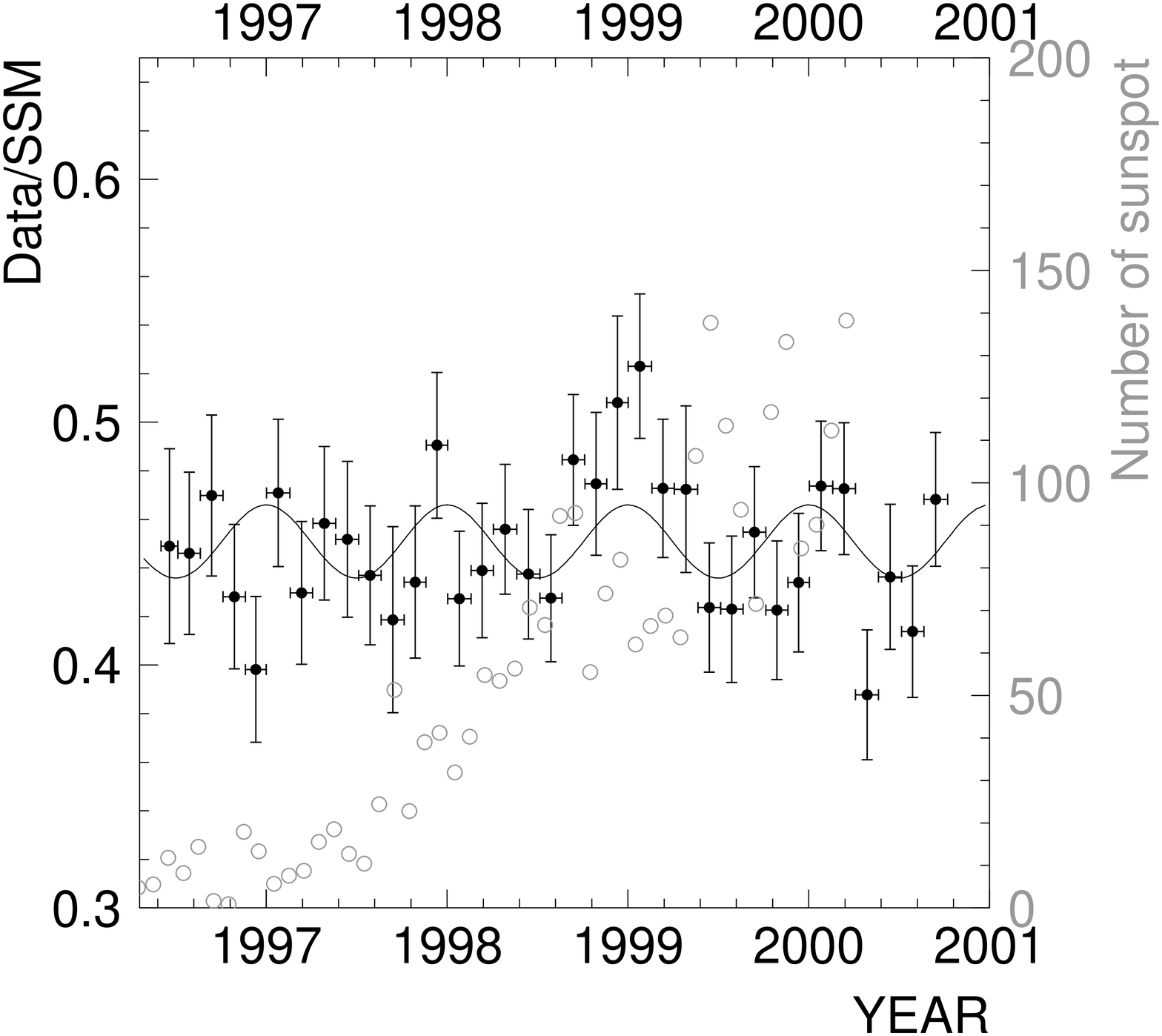,width=8.6cm}

\vspace*{-8.1cm}\hspace*{8.6cm}
\psfig{figure=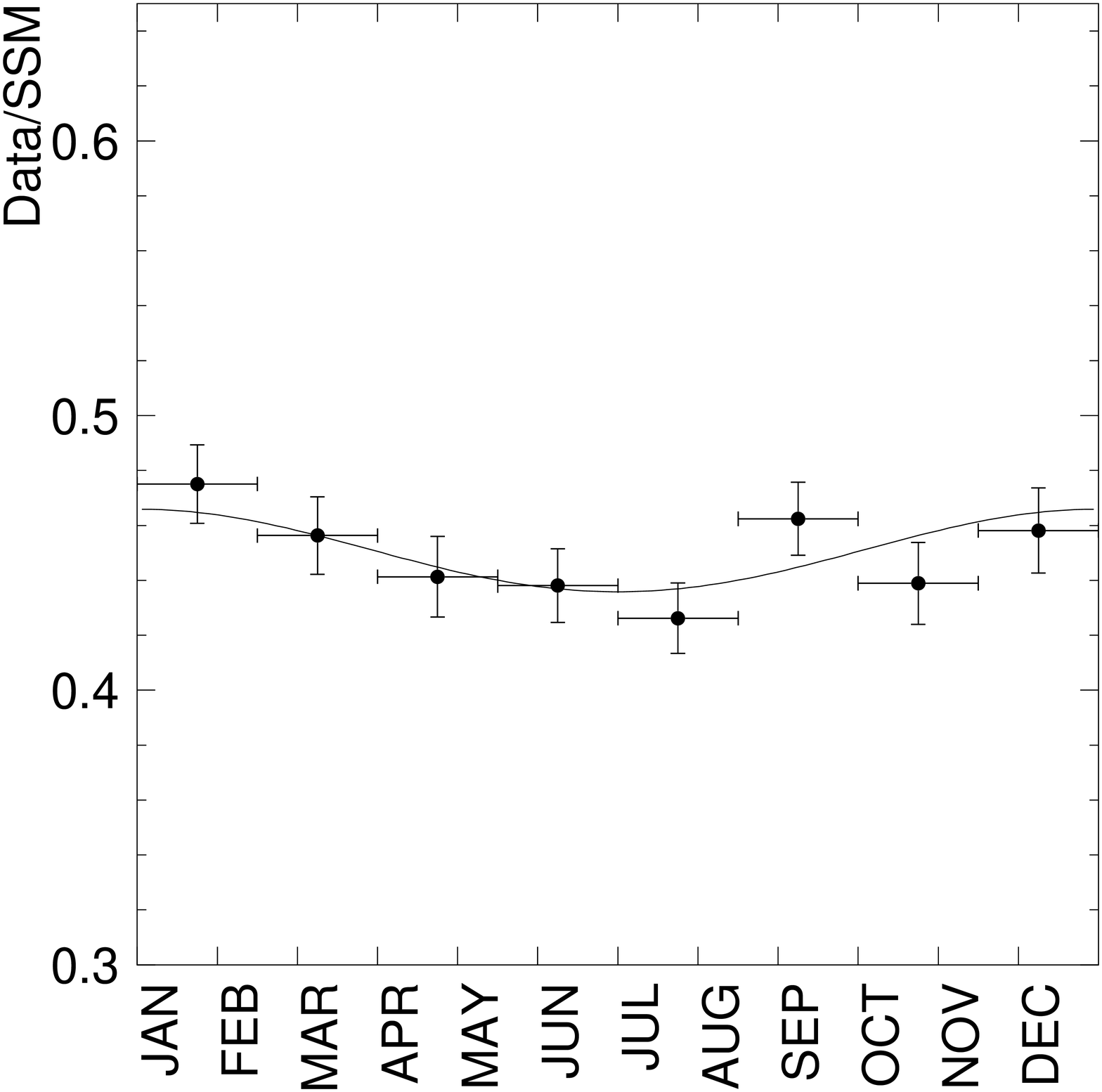,width=7.45cm}

\caption{Time-dependence of the solar neutrino flux.
Each data point extends over 1.5 months.
No correlation with sunspot activity is observed.
The solid line indicates the 7\% annual modulation
expected from the 3\% change in distance between sun
and earth. Combining the time bins into eight seasonal
bins (right) the 7\% modulation is favored by
about 2$\sigma$ over
a flat rate. 
\label{fig:timevar}}
\end{figure}

We define the zenith angle $\theta_z$ of an event as the angle between the
vertical direction and the vector sun---SK at the time of the event.
Day events have $\cos\theta_z\le0$ and night events $\cos \theta_z>0$.
Dividing the data into zenith angle bins, a flux is measured for each
sub-sample. The resulting zenith angle distribution
is shown in figure~\ref{fig:zendist}. It is consistent with a flat
suppression of about 50\% of the expected solar neutrino flux.
To test the influence of the earth's matter density on the solar
neutrino flux (a regeneration of the flux is 
predicted for some neutrino oscillation parameters)
we construct the day--night asymmetry
\be
A_{DN}=\frac{\phi_D-\phi_N}{\phi_{\mbox{av}}}
\hspace*{2cm}\mbox{where }
\phi_{\mbox{av}}=\frac{1}{2}\left(\phi_D+\phi_N\right)
\ee
from the day flux $\phi_D$ and the night flux $\phi_N$. This
asymmetry is
\be
A_{DN}=0.033\pm0.022\mbox{(stat)}^{+0.013}_{-0.012}\mbox{(syst)}.
\ee
It is consistent with zero.
Figure~\ref{fig:timevar} shows the time variation of the solar
neutrino flux. Each time bin extends over 1.5 months. No correlation
with the sunspot activity was observed. 
Due to the eccentricity of the earth's orbit,
the distance between sun and earth changes by about 3\% with
the season. As a consequence,
a 7\% annual modulation of the solar neutrino flux is expected.
Some oscillation parameters predict a 
modification of this 7\% modulation due to a change in
the oscillation phase.
To test this, the time bins are
combined into 8 seasonal bins. The seasonal distribution is
shown in figure~\ref{fig:timevar}. A $\chi^2$ test including systematic
uncertainties results in 3.8/7 degrees of freedom
(for 7\% modulation) compared to 8.1 (for flat).
Therefore,
a 7\% annular modulation is favored 
over no modulation at the 2$\sigma$ level.

\section{Neutrino Oscillation Search}

\begin{figure}[tbh]
\psfig{figure=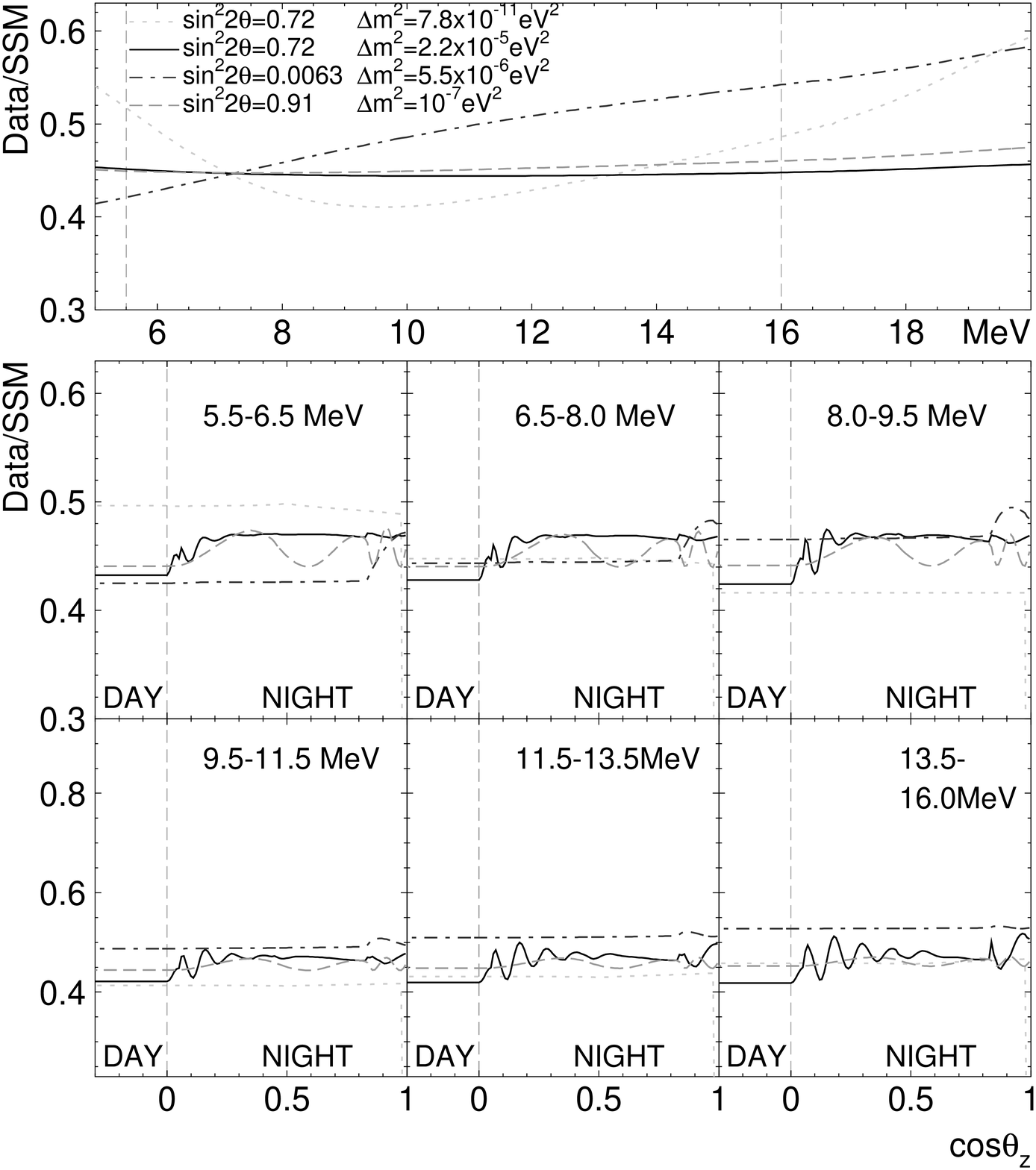,width=8cm}
\psfig{figure=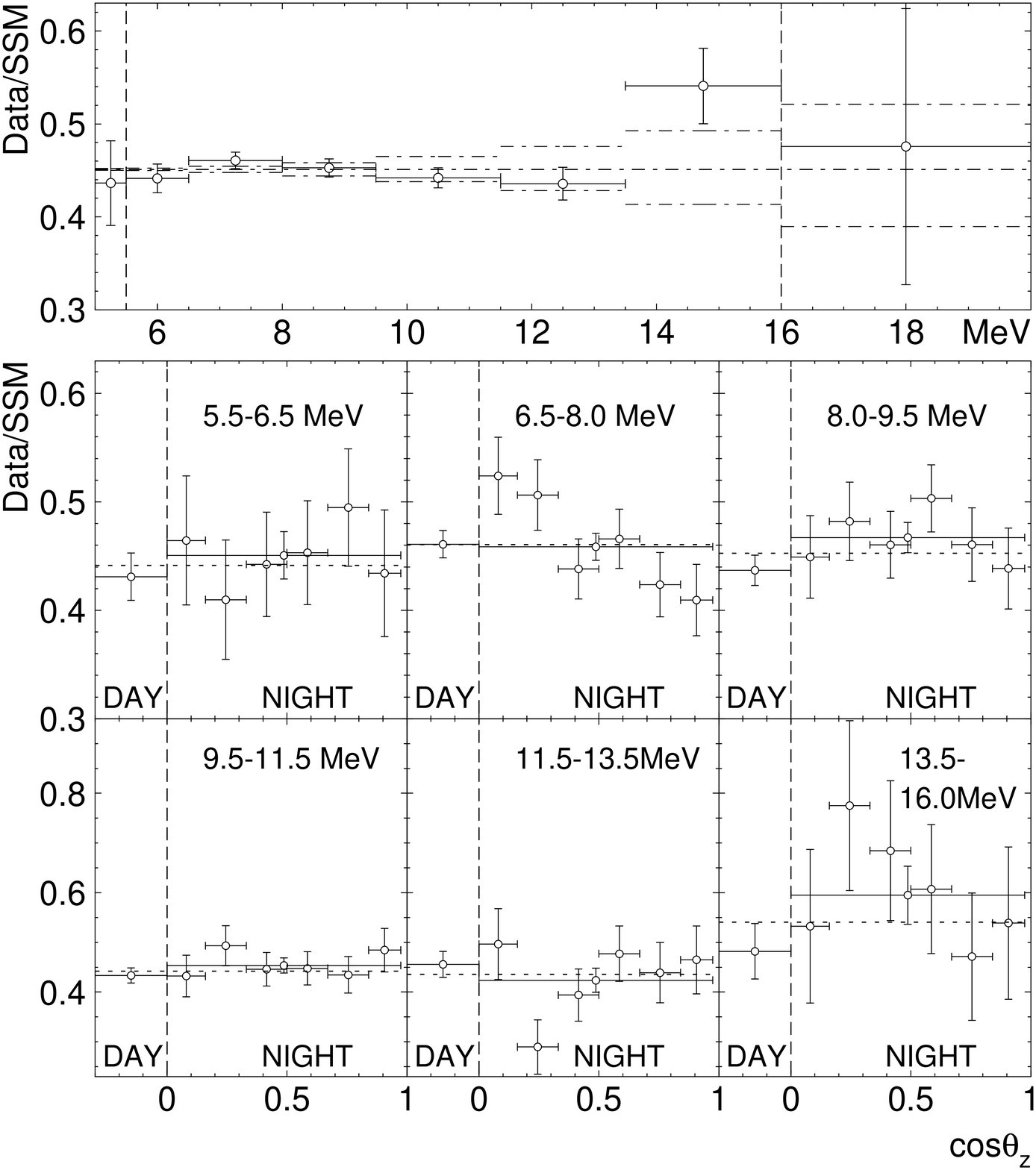,width=8cm}
\caption{Predicted (left) and observed (right)
spectral distortion (top) and zenith angle dependence (bottom)
of the solar neutrino flux. The predictions are located in the
LMA (solid), SMA (dashed-dotted), LOW (dashed) and VAC (dotted)
region.
The dashed-dotted lines in the right figure show the flux (45.1\%)
and the $\pm1\sigma$ energy-correlated systematic uncertainty.
Due to a statistical limitation,
the data sample is not broken into zenith angle bins for a
recoil electron energy below 5.5 MeV
and above 16 MeV. For the same reason, the displayed vertical axis interval
is larger above 11.5 MeV.
\label{fig:zenspec}}
\end{figure}

In addition to suppression of the solar neutrino flux or modification
of the seasonal dependence, neutrino oscillations
can distort the neutrino spectrum (and therefore the recoil electron
spectrum) or introduce zenith angle variations. 
To simultaneously study recoil electron
spectrum and zenith angle dependence of the solar neutrino flux, the data are
divided into energy bins and zenith angle bins.
SK defines eight energy bins and seven zenith angle bins
(see figure~\ref{fig:zenspec}). The
size of each energy bin is about one standard deviation of the energy
resolution function. Due to a statistical limitation, the data below
5.5 MeV and above 16 MeV are not broken into zenith angle bins. All
other data are subdivided into six (about evenly-spaced) zenith angle bins
for the night and one bin for the day. Solar neutrinos in the sixth night
bin pass through the core of the earth. This way to bin the data is
referred to as `zenith angle spectrum'.

There are several distinct two-neutrino oscillation solutions to the
solar neutrino problem: The Large Mixing Angle (LMA) solution has
a mass$^2$ difference $\Delta m^2$ of about $10^{-5}$ to $10^{-4}$eV$^2$
and a mixing angle in the range of $\sin^22\theta$=0.4 to 1. The Small
Mixing Angle Solution (SMA) has a somewhat smaller $\Delta m^2$
and mixing between $10^{-3}$ and $10^{-2}$. The mixing at the LOW solution
is close to maximal and the 
mass$^2$ difference is about $10^{-7}$eV$^2$. Below $10^{-9}$eV$^2$
there are several VAC solutions where matter effects are unimportant.
As shown in figure~\ref{fig:zenspec}, LMA solutions show little
spectral distortions, but have rapid oscillations of the flux as a
function of zenith angle. The amplitude of these oscillations is
larger for a larger recoil electron energy. Conversely, the high $\Delta m^2$
part shows almost no zenith angle variation. The LOW solution has
a small spectral distortion and slow oscillations of the flux as a
function of zenith angle. The amplitude is larger for smaller recoil
electron energy. A lower $\Delta m^2$ will then suppress these
oscillations. The SMA has moderate spectral distortion. There is almost
no zenith angle variation in the mantle region, but there is enhancement
(suppression) of the core flux at low recoil electron energy in the adiabatic
(non-adiabatic) region. The VAC solutions
typically have strong spectral distortion. The zenith angle variation
is tiny (and due to a modified seasonal flux variation in conjunction
with the correlation of season and zenith angle at
SK's geographic position).
An explanation of the MSW zenith angle signatures can be found in
\cite{earthmatter}. The measured zenith angle spectrum is most consistent
with flat, i.e. no spectral distortion and no zenith angle dependence.
It therefore favors high $\Delta m^2$ LMA and low $\Delta m^2$ LOW
solutions as well as very small mixing SMA solutions in the non-adiabatic
region, where the predicted spectral distortion is weak.

\begin{figure}[tbh]
\psfig{figure=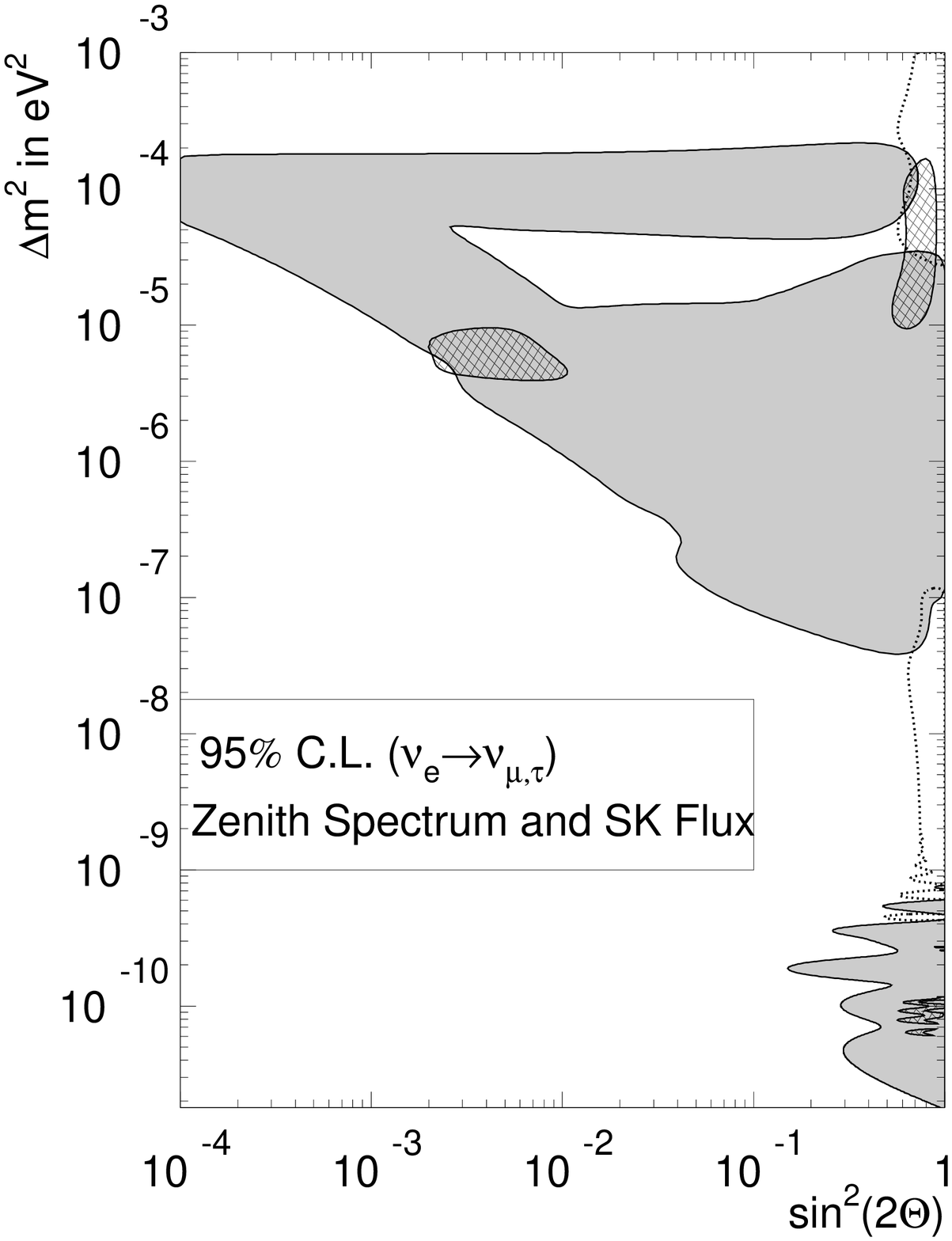,width=8cm}
\psfig{figure=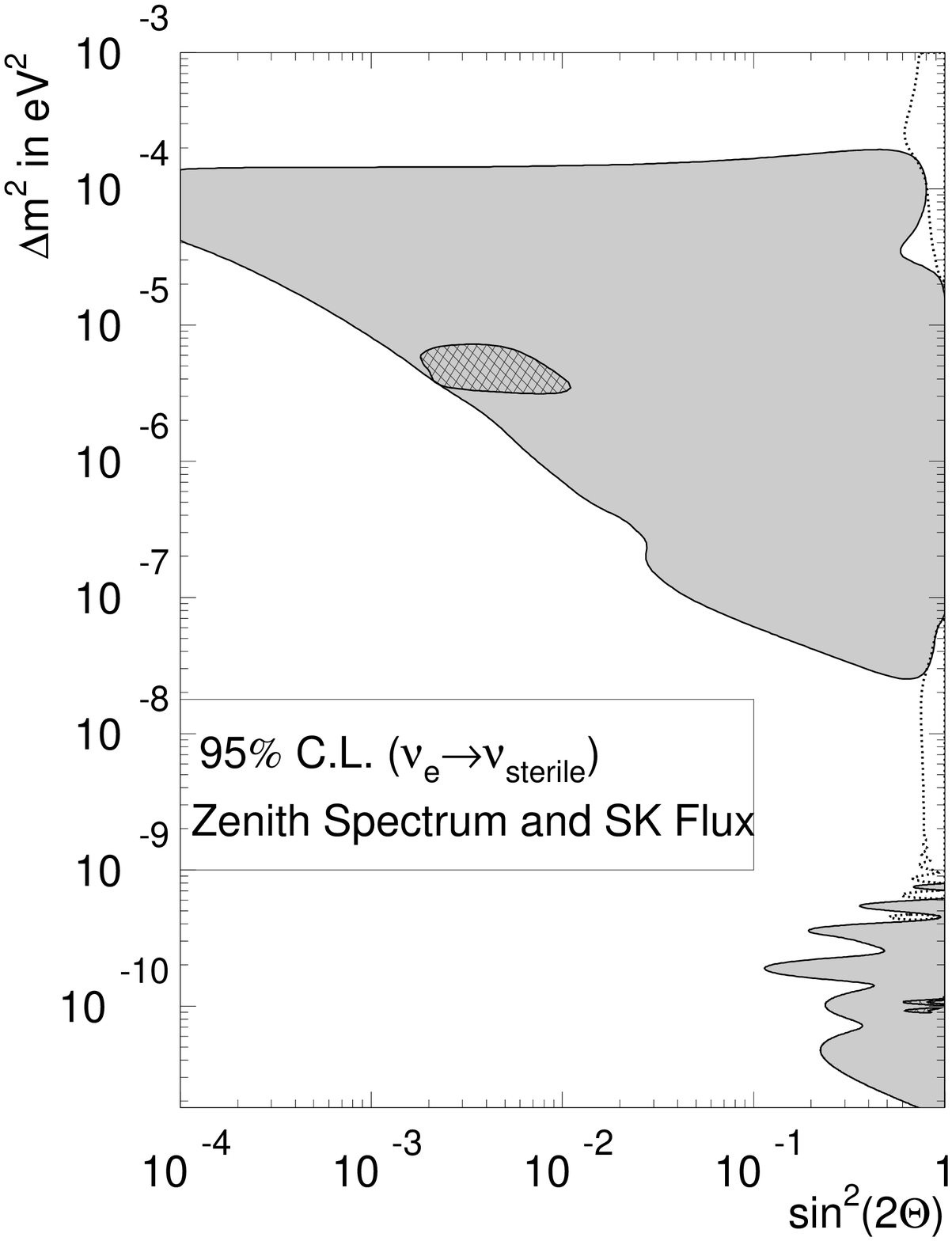,width=8cm}
\caption{Excluded (grey areas) and allowed regions (dotted lines)
for two-neutrino oscillation hypotheses. The figure on the left (right)
assumes oscillations into only active (sterile) neutrinos.
The excluded regions are based on an analysis of the zenith angle
spectrum shape alone. The allowed areas use in addition the SK rate
measurement. For comparison, the allowed areas based on
the rates of Homestake\protect\cite{homestake},
GALLEX/GNO~\protect\cite{gallex}, SAGE~\protect\cite{sage} and SK
are overlaid.
\label{fig:osc}}
\end{figure}

A calculation of the zenith angle spectrum for each oscillation parameter
was obtained as described in~\cite{skosc} for two cases:
(i) oscillation into only active ($\mu$-type or $\tau$-type) neutrinos
(ii) oscillation into only sterile neutrinos. 
As in~\cite{skosc} we define a $\chi^2$ to compare prediction against data.
A larger than expected flux from the {\it hep}
branch of solar neutrinos can mimic the spectral distortion signature
of neutrino oscillations. There is no {\it hep} flux uncertainty given
in~\cite{bp2000}. Therefore we search for the best-fit {\it hep} flux
for each oscillation parameter set. The best-fit parameters
are ($\sin^2 2\theta = 1$,
$\Delta m^2 = 6.53\cdot10^{-10}$ eV$^2$) with 36.1/40 degrees of freedom
(78.8\% normalization factor and 0 {\it hep}) in the active 
and 35.7/40 degrees of freedom (91.7\% normalization factor and 0 {\it hep})
in the sterile case. The shaded areas in figure~\ref{fig:osc} show 95\%
confidence level excluded regions based on this $\chi^2$ minimum.
Also seen in figure~\ref{fig:osc} are cross-hatched regions which
represent 95\% confidence level allowed regions (LMA, SMA and VAC)
based on the measured rates of Homestake~\cite{homestake},
GALLEX/GNO~\cite{gallex}, SAGE~\cite{sage} and SK~\cite{sk}.

\begin{figure}[tbh]
\centerline{\psfig{figure=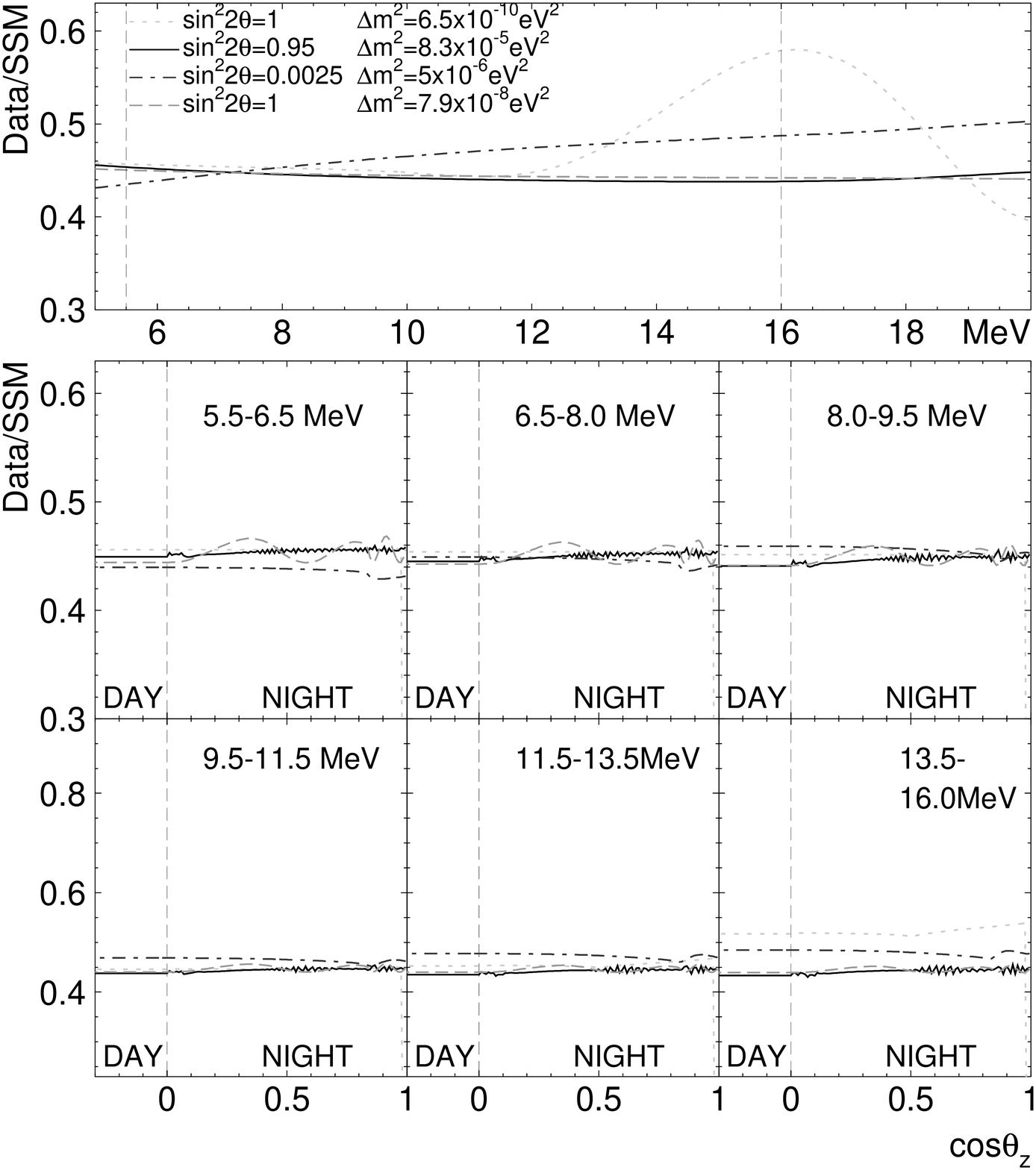,width=8cm}}
\caption{Best fit zenith angle spectrum for LMA (solid),
SMA (dashed-dotted), LOW (dashed) and VAC (dotted).
\label{fig:zenfit}}
\end{figure}

The best-fit zenith angle spectrum was searched for in these regions.
(The LOW search region was for
sin$^22\theta>0.8$ and 
$7.9\cdot10^{-8}$eV$^2<\Delta m^2<1.3\cdot10^{-7}$eV$^2$, the
VAC region for $\Delta m^2<10^{-8}$eV$^2$)
The zenith angle spectra are shown in figure~\ref{fig:zenfit}. Only
the LMA fit is truly inside the LMA region. The SMA
(in the non-adiabatic region at the lower left
corner)
and LOW best-fits are
at the boundary.
The VAC best-fit point is at higher $\Delta m^2$ than the regions shown in
figure~\ref{fig:osc}. The best-fit zenith angle spectra predict little
spectral distortion or zenith angle variation. The residual spectral distortion
of the best-fit SMA zenith angle spectrum leads to a low confidence level
(about 7\% from comparison with the $\chi^2$ minimum). 

\begin{figure}[tbh]
\psfig{figure=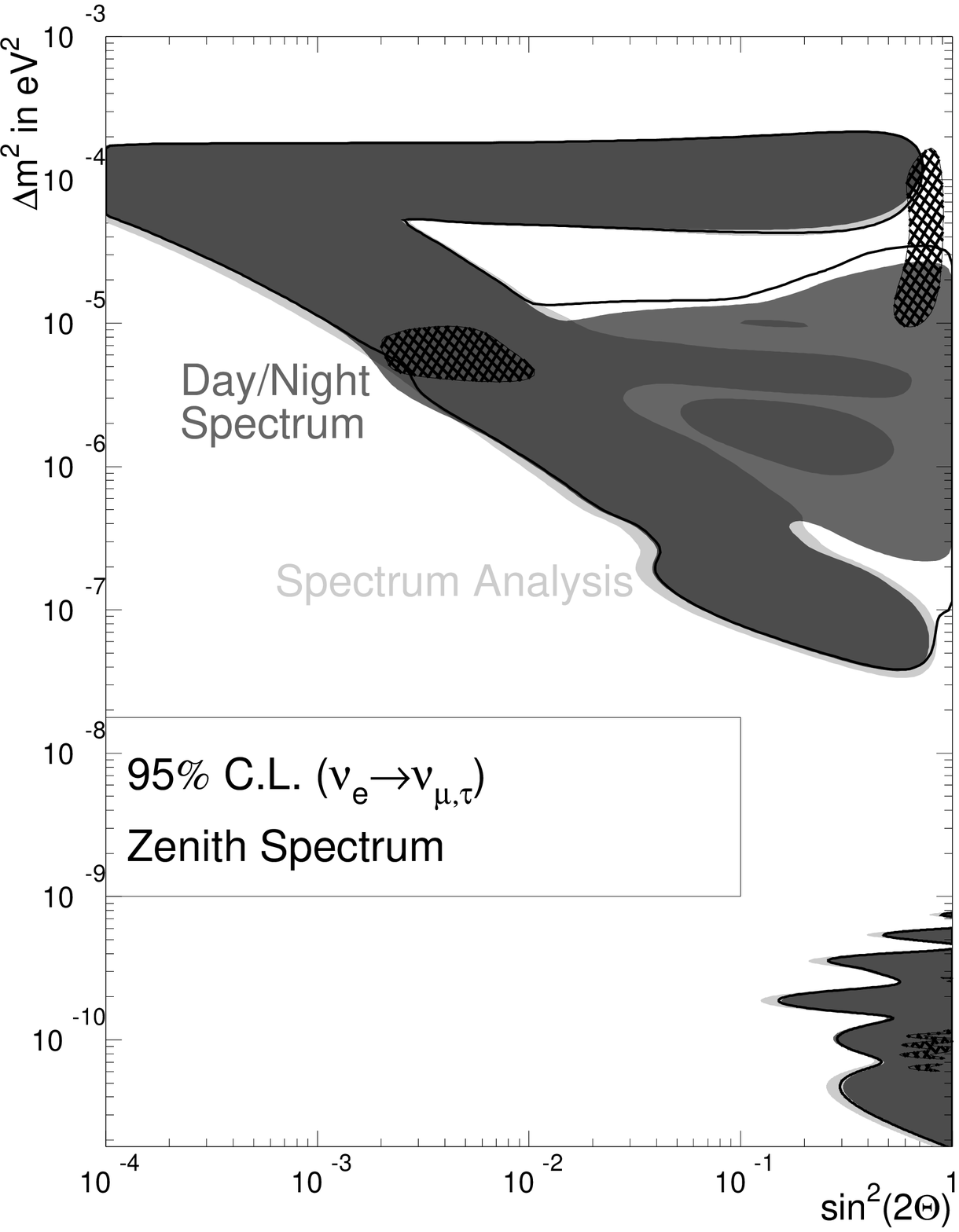,width=8cm}
\psfig{figure=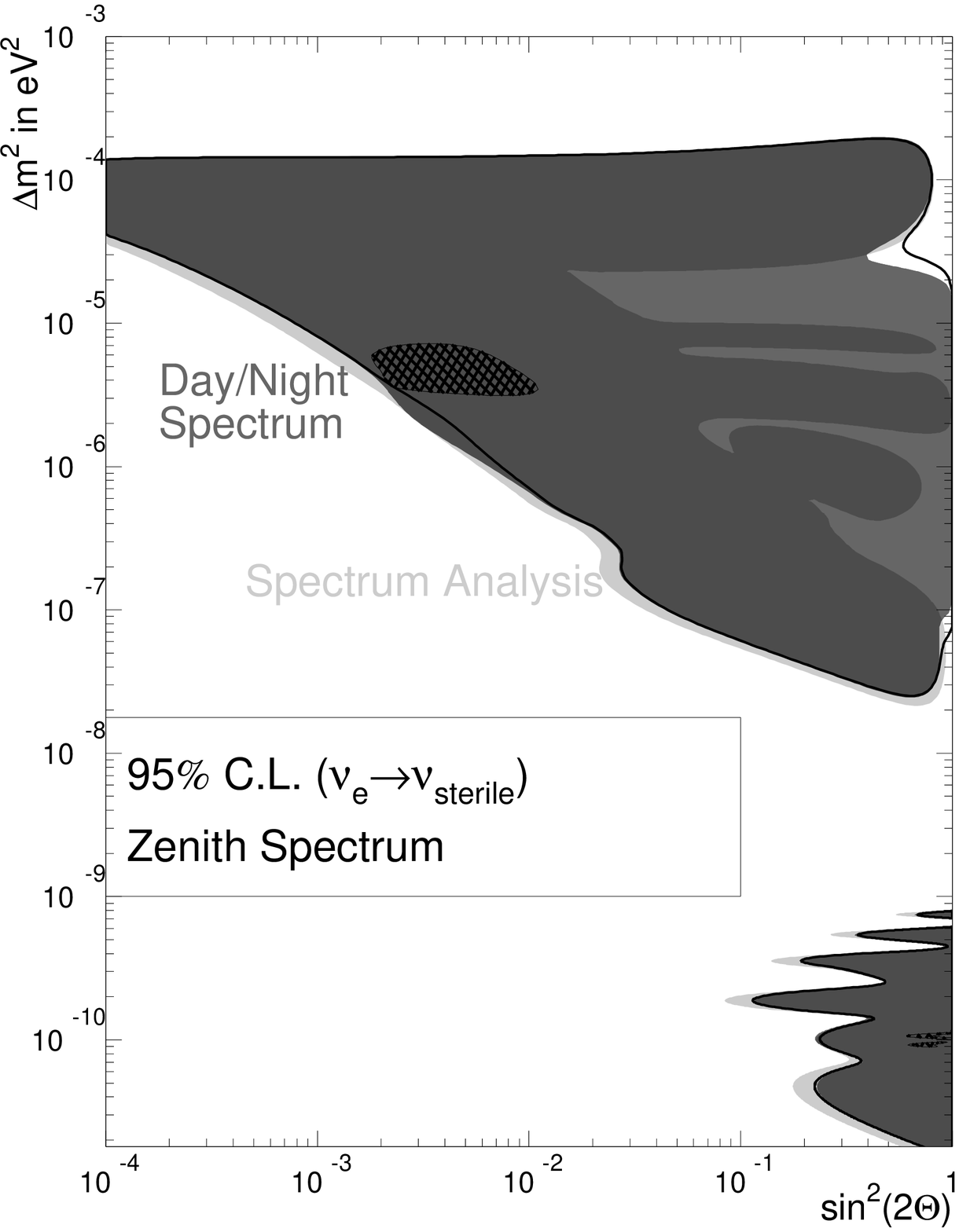,width=8cm}
\caption{Regions excluded by the shape of the day/night spectrum
(grey areas) and spectrum (light grey areas) in the active (left)
and sterile (right) case. The dark grey area is excluded by both shapes.
The inside of the solid line is excluded by the zenith angle spectrum
shape (as in figure~\protect\ref{fig:osc}).
\label{fig:osc2}}
\end{figure}

We combined the night zenith angle bin and analyzed
a day/night spectrum (eight energy bins, two zenith angle bins).
Finally, we combined all zenith angle bins to search for spectral
distortions only (eight energy bins). The resulting exclusion
regions (95\% confidence level) are shown in figure~\ref{fig:osc2}.
The day/night spectrum expands the area excluded due to spectral
distortion by a triangular region where a day--night asymmetry is
predicted. Due to the zenith angle variation inside the night bin
the zenith angle spectrum expands the excluded region further near
the LOW and the LMA
solutions. The non-adiabatic region of the SMA predicts a core
suppression. It fits better with the
zenith angle spectrum than either the spectrum or the day/night spectrum.
This is due to a slightly smaller measured flux in the core bin compared to
the day bin. Since the overall night flux exceeds the day flux,
the day/night spectrum excluded area in this region
is larger than that of the spectrum.

\begin{figure}[tbh]
\psfig{figure=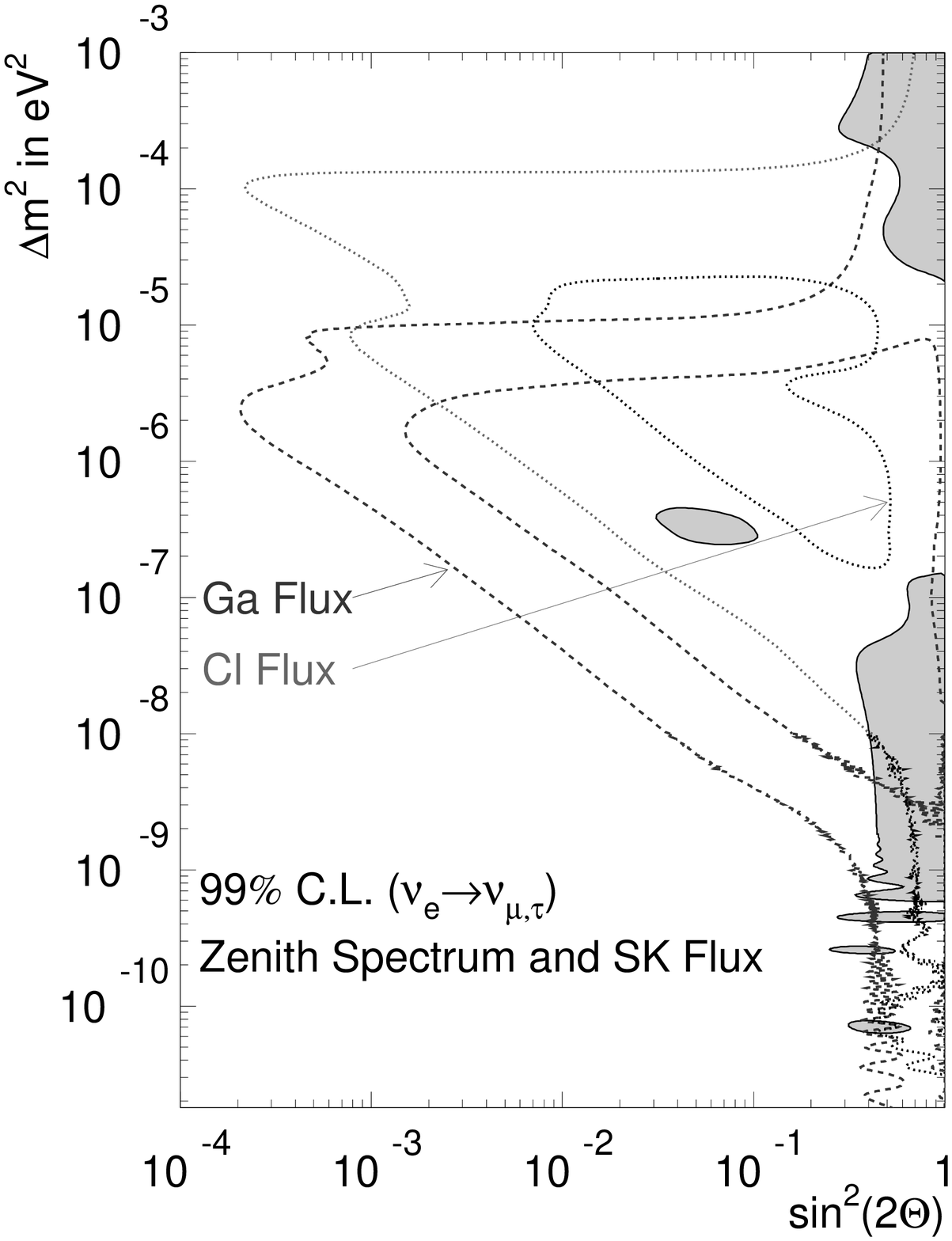,width=8cm}
\psfig{figure=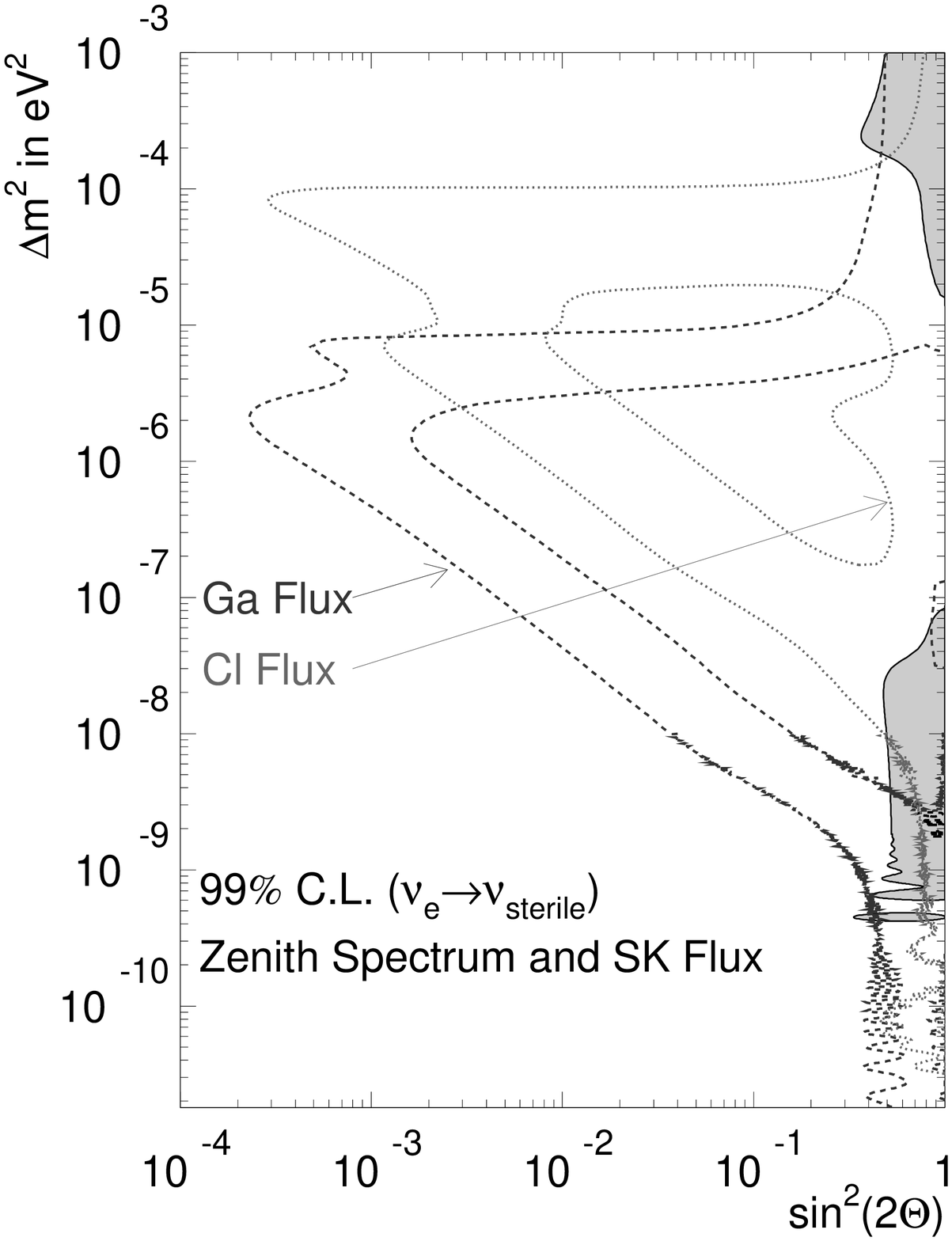,width=8cm}
\caption{99\% confidence level allowed regions using all SK data
(zenith spectrum and rate; grey areas)
compared with GALLEX/GNO~\protect\cite{gallex}+SAGE~\protect\cite{sage}
(dashed line)
and Homestake~\protect\cite{homestake} (dotted line)
for the active (left) and the sterile (right) case. 
\label{fig:osc3}}
\end{figure}

We also performed an analysis constraining the
normalization factor within the BP2000 uncertainty. There is little change
in best-fit point, normalization factor and {\it hep} flux. The
$\chi^2$ becomes 37.8/41 degrees of freedom in the active and 35.9/41
degrees of freedom in the sterile case. The region inside the dotted
lines of figure~\ref{fig:osc} is allowed at 95\% confidence level.
In both the active and the sterile case, the
95\% confidence level SMA and VAC regions are disfavored by the
zenith spectrum shape at about 95\% confidence level. The 
upper part of the LMA region is consistent with the zenith
angle spectrum shape. The LMA region
is missing in the sterile case (the sterile LMA cannot explain
the difference between Homestake's and SK's rate by either
spectral distortion or increased neutral current contribution to SK's rate).
SK data by itself (zenith spectrum and SK rate) result in
two allowed areas at large mixing for either case. These
areas are also shown in figure~\ref{fig:osc3} at 99\% confidence
level. In the active case, three small areas appear at 99\%;
one at sin$^22\theta$=0.03 to 0.1, $\Delta m^2=$2--4$\cdot10^{-7}$eV$^2$
in the MSW region and two which are close to the vacuum solution. No
allowed region appears close to the small mixing angle solution.
No new areas appear in the sterile case.

\section{Conclusion}
With 1258 days of data, Super-Kamiokande has measured the
$^8$B branch of the solar neutrino flux to be
$2.32\pm0.03$(stat)$^{+0.08}_{-0.07}$(syst)$\times10^6$/(cm$^2$s)
and the {\it hep} branch to be less than
$40\cdot10^3/($cm$^2$s)
at 90\% confidence level. No correlation of the neutrino flux
with sunspot activity has been seen. The seasonal dependence
of the neutrino flux favors the 7\% modulation due to the eccentricity
of the earth's orbit. No zenith angle dependence or distortion
of the spectrum of the solar neutrino flux has been seen; the
zenith angle spectrum is consistent with flat. In particular,
the day--night asymmetry is
$A_{DN}=0.033\pm0.022$(stat)$^{+0.013}_{-0.012}$(syst).
The absence of zenith angle dependence and spectral distortion
strongly constrains two-neutrino oscillations. From the six
solution areas based on the rates of Homestake, GALLEX/GNO, SAGE
and SK 
--- active VAC, active SMA, active LOW, active LMA, sterile VAC
and sterile SMA ---
all but active LMA are disfavored by the shape of the
zenith angle spectrum at about 95\% confidence level. Using only
SK data (zenith angle spectrum shape and SK rate), two allowed
areas (95\% confidence level) at large mixing are found for the active as well
as the sterile case: $\Delta m^2>2$--$3\cdot 10^{-5}$eV$^2$
or $10^{-9}$eV$^2<\Delta m^2<10^{-7}$eV$^2$.

\section*{Acknowledgments}
The authors acknowledge the cooperation of the Kamioka Mining and
Smelting Company.  The Super-Kamiokande detector has been built and
operated from funding by the Japanese Ministry of Education, Culture,
Sports, Science and Technology, the U.S. Department of Energy, and the
U.S. National Science Foundation.  This work was partially supported
by the Korean Research Foundation (BK21) and the Korea Ministry of
Science and Technology.

\end{document}